\documentstyle[12pt]{article}

\begin{document}
%\begin{center}
{\center{\bf{

\Huge{Statistical Shapes of the Highest Pulses in Gamma--Ray Bursts\footnote{Supported by the Special Funds
for Major State Basic Research Projects, the National Natural Science Foundation of China, and the Natural
Science Foundation of Yunnan province, China.}}
%%% begin:list of authors

}}} \vspace{20pt}

\small{Yi--Ping Qin$^{1}$, En--Wei Liang$^{1,2}$, Guang-Zhong Xie$^{1}$, and Cheng-Yue Su$^{1,3}$
$^{1}$Yunnan Observatory, National Observatories, Chinese Academy of Sciences, Kunming 650011, P. R. China,
email:qinyp@public.km.yn.cn \\ $^{2}$Physics Department, Guangxi University, Nanning, Guangxi 530004, P. R.
China,\\ $^{3}$Department of Physics, Guangdong Industry University, Guangzhou 510643, P. R.
            China}
%\end{center}

\begin{abstract}

The statistical shapes of the highest pulse have been studied by
aligned method. A wavelet package analysis technique and a
developed pulse--finding algorithm have been applied to select the
highest pulse from burst profiles observed by BATSE on board CGRO
from 1991 April 21 to 1999 January 26. The results of this work
show that the statistical shapes of the highest pulses are related
to energy: the higher the energy, the narrower the pulse. However,
the characteristic structures of the pulses have nothing to do
with energy, which strongly supports the previous conclusion that
the temporal profiles in different channels are self--similar. The
characteristic structures of the pulses can be well described by a
model proposed by  Norris et al. (1996). The fitting parameters
are: $t_r$=0.12, $t_d$=0.16, $\upsilon =1.09$, the ratio of $t_r$
to $t_d$ for the pulse is 0.75. The result leads to our conjecture
that the mechanisms of bursts in different gamma-ray bands might
be the same. The shock, either an internal or an external one,
producing the pulse, might emit photons over the four energy
channels in the same way.

\end{abstract}
{\center{{\bf{gamma rays: bursts --- methods: data analysis}}}}
\section{Introduction}

Gamma--ray bursts(GRBs), which are still mysterious, have very
complex temporal structure. Their temporal profiles are enormously
varied --- no two bursts have ever been found to have exactly the
same temporal and spectral
development. The temporal activity is suggestive of a stochastic process (%
Nemiroff et al. 1993). The diversity of the bursts seems to be due to random realization of the same process
that is self-similar over the whole range of timescale. Attempts to quantify these structures have not been
successful (e.g., Fishman 1999).

Most of the observed profiles of GRBs are composed of pulses, each comprising a fast rise and an exponential
decay (a FRED; e. g., Desai 1981, Fishman et al. 1994). Many methods for pulse analysis have been developed,
e. g., the parametric analysis in model fitting (Nemiroff et al. 1993, Norris et al. 1996), the
auto-correlation method (Fenimore et al. 1995), the nonparametric method (Li \& Fenimore 1996%
), the peak alignment and normalized flux averaging method (Mitrofanov et al. 1996, Mitrofanov et al. 1998,
Ramirez--Ruiz \& Fenimore 1999, Ramirez--Ruiz \& Fenimore 2000), and the pulse decomposition analysis method
(Lee et al. 2000), etc. These statistical studies have revealed many observed temporal signatures of pulses.
The pulses are hypothesized to have the same shape at all energies, differing only by scale factors in time
and amplitude (``pulse scale conjecture''). And, the pulses are hypothesized to start at the same time,
independent of energy (``pulse start conjecture''). The two conjectures were confirmed by Nemiroff (2000). In
general, higher energy channels show shorter temporal scale factors (e.g., Norris et al. 1996., Nemiroff
2000). It is found that the temporal scale factors between a pulse measured at different energies are related
to that energy by a power law, possibly indicating a simple relativistic mechanism is at work (Fenimore et
al. 1995, Norris et al. 1996., Nemiroff 2000).

The statistical pulse shape has been well studied by the peak
alignment and normalized flux averaging method. The peak aligned
averaging pulse is spiky. A succinct pulse model, which well
describes many pulse shapes, was proposed by Norris et al. (1996):
\begin{equation}
I(t)=I_0e^{-(|t-t_{\max }|/{t_{r,d}})^\upsilon }
\end{equation}
where $t_{max}$ is the time of the pulse's maximum intensity
($I_0$); $t_r$ and $t_d$ are the rise and decay time constants,
respectively; and $\upsilon $ is a measure of the pulse sharpness,
which was referred to ``peakness'' by Norris et al. (1996).

However, both the duration and total count of pulses vary
significantly. Statistical properties of the pulses revealed by
the peak alignment and flux--normalized averaging method are
limited. In this paper, this method is developed to study the
shape of pulses in a more detailed manner.

To reach a result of high quality, we concern in this paper only the highest pulse of bursts, where one finds
the highest level of signal--to--noise. We make the noise decomposition for the time profile of bursts by
performing the wavelet analysis (which is described in section 2), then modify the pulse--finding algorithm
proposed by Li \& Fenimore (Li \& Fenimore 1996) to identify the highest pulse in a burst profile (see
section 3). In sections 4 and 5, we employ and develop the pulse aligned method to study the flux--normalized
aligned averaging pulse shape and the count--and--duration--normalized aligned averaging pulse shape,
respectively. Conclusions and discussion are presented in section 6.

\section{Data Analysis}

The data used for analyzing is the 64 ms temporal resolution and
four--channel spectral resolution GRB data observed by BATSE from
1991 April 21 to 1999 January 26. There are 1738 bursts included.
It is a concatenation of three standard BATSE data types, DISCLA,
PREB, and DISCSC. All these data types are derived from the
on-board data stream of BATSE's eight Large Area Detectors (LADs).
There are four energy channels observed, with the following
approximate channel boundaries: 25-55 keV, 55-110 keV, 110-320
keV, and $>$320 keV. The DISCLA data are a continuous stream with
1.024 second resolution. They are independent of burst occurrence
and taken as the background. The PREB data cover the interval
2.048 second just prior to a burst trigger.

We make the noise decomposition for the time profiles by the wavelet package analysis technique. The
technique is suitable to treat those signals which cannot be analyzed by the traditional Fourier method. It
was successful in de--noising the original signal and identifying the structure within a burst (e.g., Hurley
et al. 1997, Quilligan et al. 1999, Lee et al. 2000). We use DB3 wavelet to make the first--class
decomposition with the MATLAB software. The profile is decomposed into the signal component and the noise
component. Figure 1 illustrates an example of the decomposition.

The method of the background treatment used here is similar to
that in Li \& Fenimore (1996). Since the DISCLA data are a
continuous stream prior to and independent of the burst
occurrence, they are always taken as the background of bursts. The
data of the background is obtained by a linear fitting to the
DISCLA data.

\begin{figure}[tbp]
\vbox to 3.0in{\rule{0pt}{3.0in}}

\includegraphics{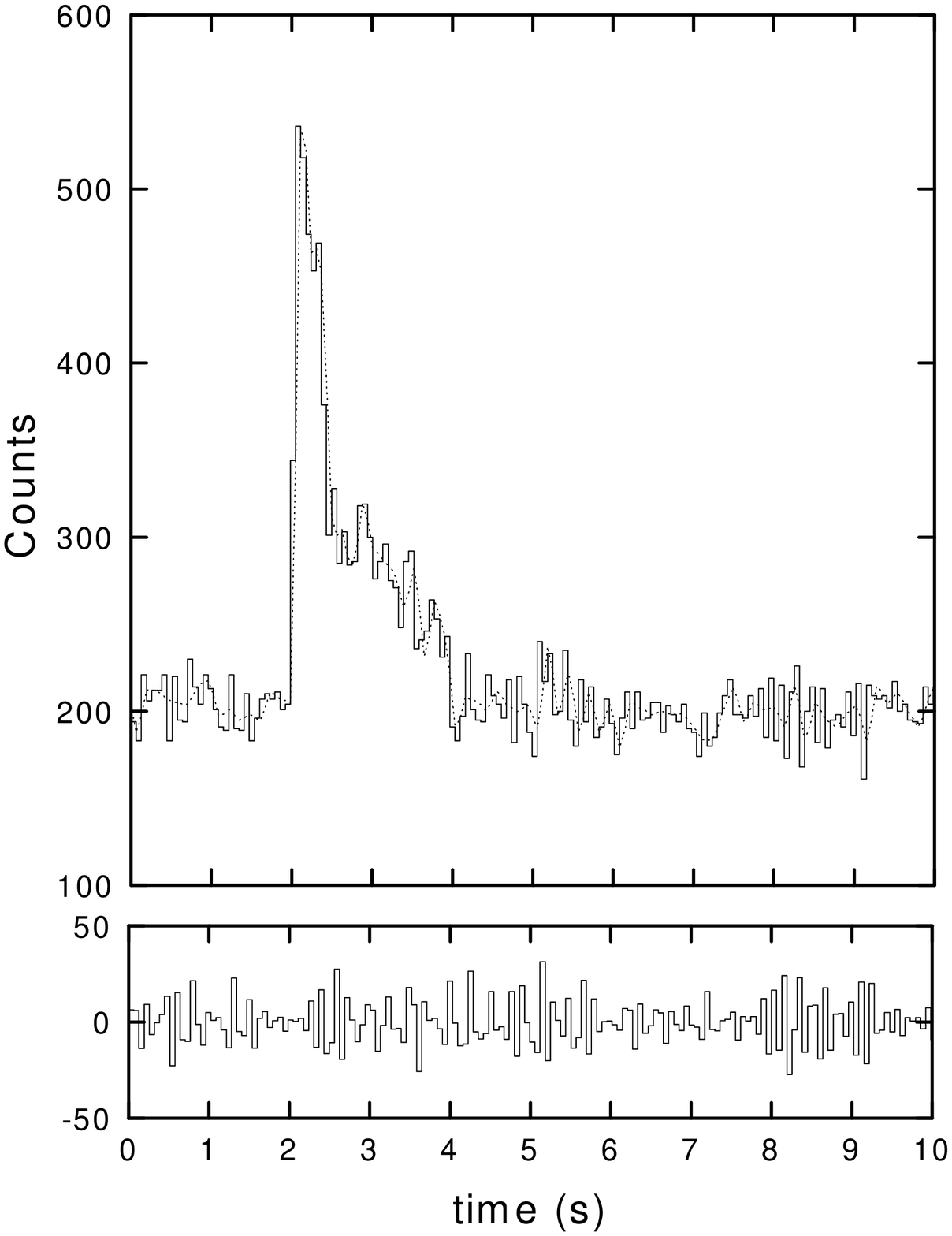} \caption{Illustration of the
noise decomposition using the wavelet package analysis technique for trigger 6341 in channel 3. Presented in
the upper panel are the original data (solid step line) and the signal component (dotted straight line),
while the noise component is displayed in the lower panel. } \label{fig1}

\end{figure}

\section{Pulse--Finding Algorithm and Sample Selection}

Many burst time profiles appear to be composed of a series of overlapping pulses, mingling with noises. It is
not easy to determine their actual light curves and to distinguish a pulse from the time profile. The result
of pulse analysis strongly relies on the algorithm of pulse--finding and sample selection. Several
pulse--finding algorithms have been proposed (e.g., \cite {LF96}, Norris et al. 1996,Mitrofanov et al. 1998).
Li \& Fenimore (1996) suggested an efficacious algorithm to identify ``true peaks'' in a profile. A ``true
peak'' is not necessarily to be regarded as a pulse. If a profile is composed of only one ``true peak'', the
``true peak'' can be regarded as a pulse. However, most of the profiles are composed of many overlapping
``true peaks''. To identify a pulse in such situation is not easy. Norris et al. (1996) introduced a
definition of ``inseparable pulse''. We adopt this concept and regard an ``inseparable pulse'' as a true
pulse. Our pulse--finding algorithm is described in the following.

(1) The peak--finding criterion proposed by Li \& Fenimore is $%
C_p-C_{1,2}\geq N_{var}\sqrt{C_p}$ , where $C_i$ ($i=1,2$) is the
photon count at time bin $t_i$, $C_p$ (at $t_p$) is the maximum
count of a candidate peak, and $N_{var}$ is an adjustable
parameter, typically $3\leq N_{var}\leq 5$. This criterion
strongly relies on absolute photon count of the candidate peak. We
follow Norris et al. (1996) to employ the concept of
``inseparable pulse'', and then the pulse--finding criterion becomes $%
1-C_{1,2}/C_p\geq 0.5$. It means that a candidate peak is a true
peak only
when $C_1$ (at $t_1$) and $C_2$ (at $t_2$) are lower than the half of the $C%
_p$ on both sides of $t_p$. With this method, one might find more
than one true peaks within a burst. For the reason mentioned in
section 1, we select only the highest one.

(2) In order to maintain a high level of signal--to--noise, we
adopt the intensity criterion as $C_{max}>10\sigma $, where
$\sigma $ is the standard deviation of the background.

(3) Only those pulses with at least 10 bins of time are selected.
Those with less bins do not provide enough structure information
and thus are ignored.

We apply the above pulse--finding algorithm to select highest
pulses in the profiles of bursts. There are 760, 885, 885, and 334
bursts, for which the highest pulses can be identified, in
channels 1 to 4, respectively. The number of bursts for which all
of the highest pulses in four channels can be
identified is 275. The flux in this sample ranges widely, from 0.513 photons.cm$%
^{-1}$.s$^{-1}$ to 183.370 photons.cm$^{-1}$.s$^{-1}$. We select
the sample to study the statistical properties of the pulse
morphology.

\section{The Flux--Normalized aligned Averaging Pulse}

It was found that the peak--aligned averaging pulse well illustrates how the average pulse evolves with
energy. To do that, one averages the time profile of individual events by the normalized peak--alignment
technique, where each time profile is normalized by the peak number of counts C$_{max}$, aligned at the peak
time bin t$_{max}$, and then averaged for all bins along the timescale(e. g., Norris et al. 1996, Mitrofanov
et al. 1996, Ramirez--Ruiz \& Fenimore 2000). Though the pulse--finding method and the sample adopted in this
paper are somewhat different from the previous ones, the peak--aligned averaging pulses we obtain are quite
similar to that in Norris et al. (1996) and Ramirez-Ruiz \& Fenimore (2000) (see Fig. 2). Figure 2 shows the
same result that the higher the energy, the narrower the pulse. One can also find this from the
flux--normalized--and--beginning--aligned averaging pulses shown in Fig. 3.

We find that the shapes of pulses in Fig. 2 are quite different
from that in Fig. 3. Though both figures come from the sum of the
same normalized pulses, but the ways of the alignment are
different. The difference between Figs. 2 and 3 must come from the
diversity of the duration and the asymmetry of the shape of the
normalized pulses. Obviously, the peak--aligned method would lead
to a spiky shape. The statistical result must conceal most of the
diversity of the duration and the asymmetry of the pulses. This
explains why the averaging pulses in Fig. 2 are very different
from observation. Also, the flux--normalized--beginning--aligned
method does not take into account the diversity of the duration of
pulses. Thus, neither Fig. 2 or 3 truly embodies the temporal
structure of pulses.

\begin{figure}[tbp]
\vbox to 3.0in{\rule{0pt}{3.0in}} \includegraphics{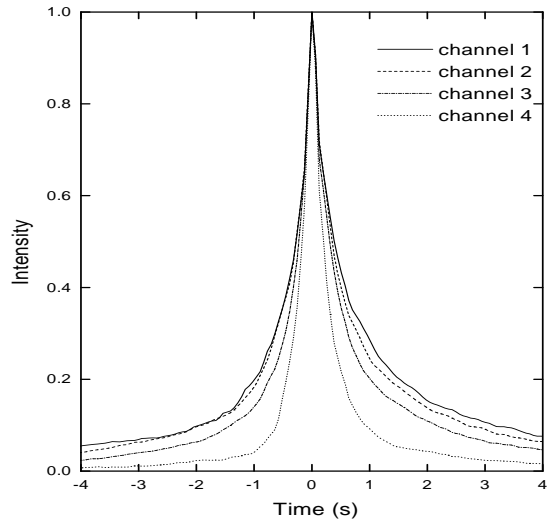} \caption{ The flux--normalized--peak--aligned averaging pulses.} \label{fig8}
\end{figure}

\begin{figure}[tbp]
\vbox to 3.0in{\rule{0pt}{3.0in}} \includegraphics{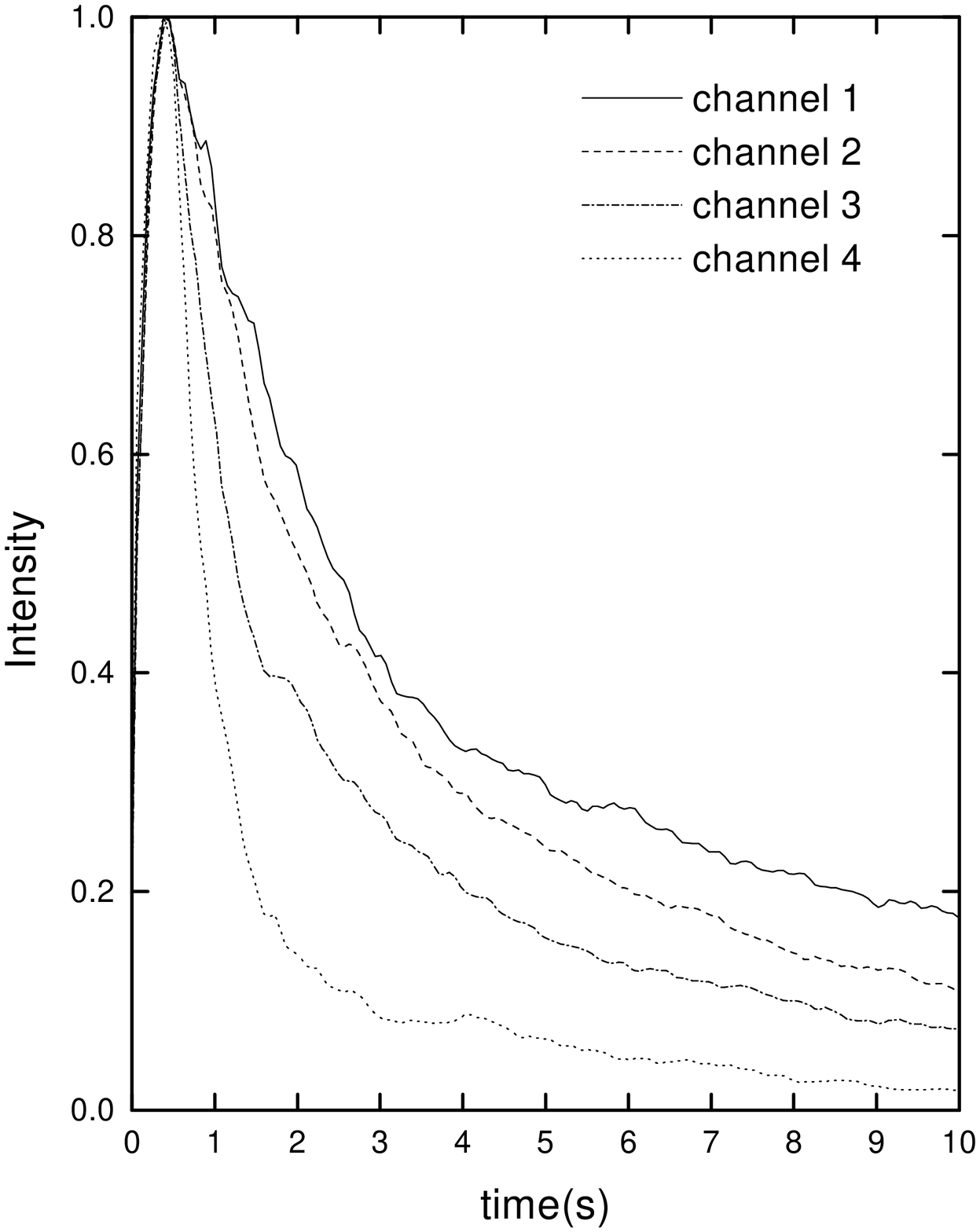} \caption{ The flux--normalized--beginning--aligned averaging pulses.} \label{fig9}
\end{figure}

\section{The Count-and-Duration-Normalized aligned Averaging Pulse}

We observe that, the above peak--aligned method allows those
pulses with longer durations to possess larger total counts and
then to contribute more significantly to the averaging pulse. To
investigate the averaged shape of pulses, one needs to get rid of
the effects from both the total count and the duration of the
selected pulses. This leads to the count-and-duration-normalized
aligned averaging method employed bellow.

First, both the total count and the duration for each selected
pulse are normalized. Then the averaging pulses for the four
channels are obtained with the same way used in last section. The
results are shown in Figs. 4 and 5. They are different
respectively from that in Figs. 2 and 3.

\begin{figure}[tbp]
\vbox to 3.0in{\rule{0pt}{3.0in}} \includegraphics{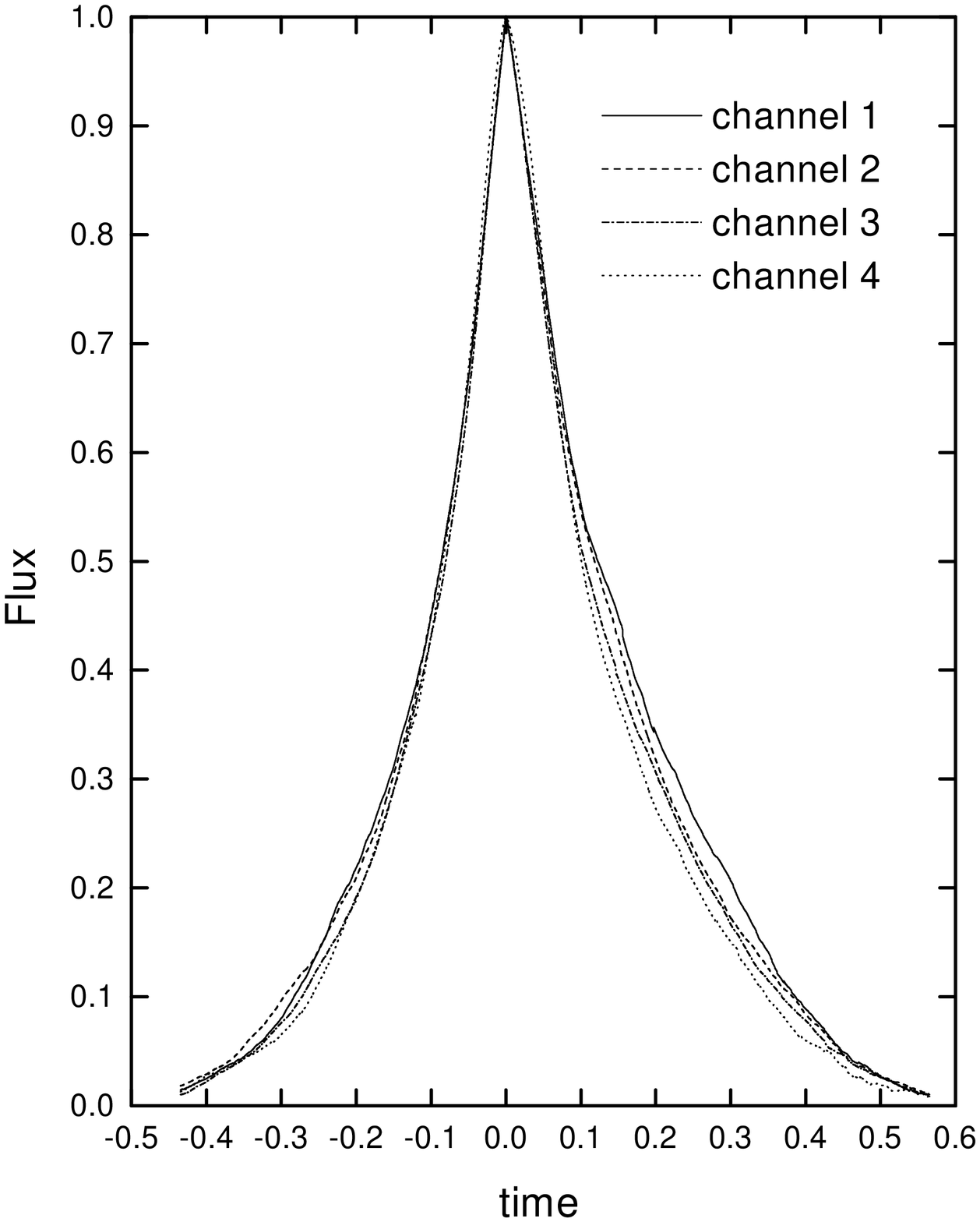} \caption{The count--and--duration--normalized--peak--aligned averaging pulses.} \label{fig10}
\end{figure}

\begin{figure}[tbp]
\vbox to 3.0in{\rule{0pt}{3.0in}} \includegraphics{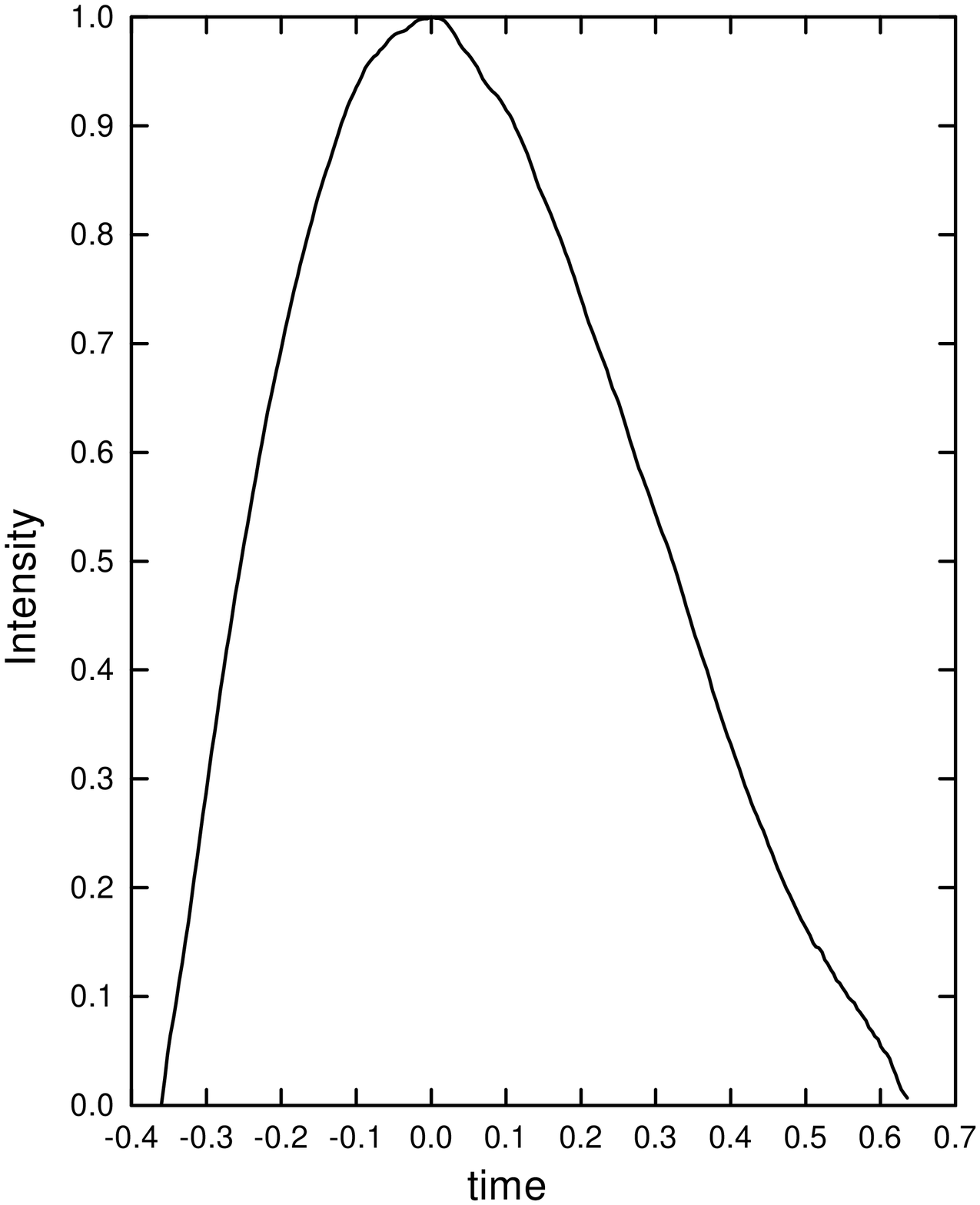} \caption{The count--and--duration--normalized--beginning--aligned averaging pulses.} \label{fig11}
\end{figure}

Figure 4 shows that the
count--and--duration--normalized--peak--aligned averaging pulses
in the four channels are almost the same. One can not tell any
difference between the
count--and--duration--normalized--beginning--aligned averaging
pulses of the four channels. It shows that the shapes of these
pulses are independent of energy bands. The pulses of the four
channels have the same temporal structure.

The pulse model in Norris et al. (1996) is employed to fit them,
and it shows that these pulses can be well described by the model.
The fitting parameters are: $t_r$=0.28, $t_d$=0.38, $\upsilon
=2.29$ for the count--and--duration--normalized--peak--aligned
averaging pulse (in channel 3); $t_r$=0.12, $t_d$=0.16, $\upsilon
=1.09$ for the
count--and--duration--normalized--beginning--aligned averaging
pulse (in channel 3). The ratios of $t_r$ to $t_d$ for both pulses
are about 0.75.

\section{Conclusions and Discussion}

In this paper we study the
flux--normalized--peak(beginning)--aligned averaging pulses and
the count-and-duration-normalized--peak(beginning)--aligned
averaging pulses with the highest pulse in the profiles of GRBs.
We apply the wavelet package analysis technique and a developed
pulse--finding algorithm to select the highest pulse. The sample
we get includes 275 bursts which fluxes range from 0.513
erg.cm$^{-1}$.s$^{-1}$ to 183.370 photons.cm$^{-1}$.s$^{-1}$.

The wavelet package analysis technique is suitable to treat those
signals which cannot be analyzed by the traditional Fourier
method. It is successful in de--noising the original signal and
identifying the structure within a burst. The pulse--finding
algorithm used in Li \& Fenimore (1996) is developed in this paper
so that the selection of pulses depends on the relative value of
counts rather than the absolute value.

The number of bursts concerned in this paper is the largest one of that used for pulse analysis so far, and
the sample adopted here covers the biggest flux range. Many samples of very bright bursts have been employed
to study statistical properties of pulses (e.g., Norris et al. 1996). Though the numbers of bursts concerned
are much smaller, the authors were able to get more pulses by selecting not only the highest pulse but also
other pulses in a burst (e.g., Norris et al. 1996). Their results refer only to bright bursts, and the pulses
so selected might include possible evolutionary effects of pulses.

Figures 2 and 3 support the well-known conclusion that the higher
the energy, the narrower the pulse.

Different from the previous works, we make in this paper not only
the normalization of the total count but also the normalization of
the duration. In this way, all the pulses (strong or weak)
contribute equally to the averaging pulses. And, in this way, the
averaging pulses obtained stand only for the statistical shape of
pulses. The effects from both the duration and the total count are
removed. The results so obtained are quite different from that got
by the previous method. One can find this by comparing Figs. 2 and
4.

For the pulses shown in Figs. 4 and 5, we prefer those in the
latter. Since the pulses in Fig. 4 come from aligning the
normalized pulses at the moment of maximum count of the pulses,
those asymmetric normalized pulses would contribute differently to
different sides of the averaging pulses. Thus the distribution of
the peak count position in the shape of selected pulses must be at
work. It would lead to a spiky shape of the averaging pulses. Fig.
4 must conceal most of the diversities of the duration and the
asymmetry of pulses. Differently, the pulses shown in Fig. 5 stand
only for the average shape of the original pulses. These pulses
are well described by the model in Norris et al. (1996). The
fitting
parameters for the pulse in channel 3 in Fig. 5 are: $t%
_r$=0.12, $t_d$=0.16, $\upsilon =1.09$. The ratios of $t_r$ to
$t_d$ for the pulse is 0.75.

We find that the
count-and-duration-normalized--peak(beginning)--aligned averaging
pulses are the same for different channels. Our results strongly
support the previous conclusion that the temporal profiles in
different channels are self--similar. The averaging pulse shape is
independent of energy bands. Due to these results, we believe that
the mechanisms of bursts in different gamma-ray bands must be the
same.

The mechanism generating the bursts is still unknown. Many models for interpreting the origin and emission of
the event have been proposed (e. g., Rees \& M$\acute{e}$sz$\acute{a}$ros 1992, Vietri et al. 1998, Fuller et
al. 1998, Dai \& Lu 1998, Daigne et al. 1998, \cite {Pa00}, etc.), mostly in the context of two major
scenarios involving relativistic shells. An approach frequently used in these models is to identify each
pulse in the light curve with a single event. Depending on the model chosen, this event could be the
collision between inhomogeneities in a relativistic wind in the internal models or the ``activation'' of a
region on a single external shell. Our study shows that, the shock, either an internal or an external one,
producing the pulse, might produce photons over the four energy channels in the same way.

\clearpage

\end{document}